\documentclass[amsmath,amssymb,aps,twocolumn]{revtex4-1}

\usepackage{amssymb}
\usepackage{graphicx}
\usepackage{dcolumn}
\usepackage{bm}
\usepackage{amsmath}
\usepackage{mathrsfs}
\usepackage{textcomp}
\usepackage{upgreek}
\usepackage[unicode=true,bookmarks=true,bookmarksnumbered=false,bookmarksopen=false,breaklinks=false,pdfborder={0 0 1},backref=false,colorlinks=true]{hyperref}

\hypersetup{linkcolor=magenta,urlcolor=blue,citecolor=blue,pdfstartview={FitH},hyperfootnotes=false,unicode=true}

\def\be{\begin{equation}}
	\def\ee{\end{equation}}
\def\bea{\begin{eqnarray}}
	\def\eea{\end{eqnarray}}

\begin{document}

\preprint{APS/123-QED}
\title{Observation of the antiferromagnetic phase transition in the fermionic Hubbard model}
\author{Hou-Ji Shao$^{1,2}$}
\thanks {These authors contributed equally to this work.}
\author{Yu-Xuan Wang$^{1,2}$}
\thanks {These authors contributed equally to this work.}
\author{De-Zhi Zhu$^{1,2}$}
\author{Yan-Song Zhu$^{1,2}$}
\author{Hao-Nan Sun$^{1,2}$}
\author{Si-Yuan Chen$^{1,2}$}
\author{Chi Zhang$^{1,2}$}
\author{Zhi-Jie Fan$^{1,2,3}$}
\author{Youjin Deng$^{1,2,3}$}
\author{Xing-Can Yao$^{1,2,3}$}
\author{Yu-Ao Chen$^{1,2,3}$}
\author{Jian-Wei Pan$^{1,2,3}$}
\affiliation{$^1$Hefei National Research Center for Physical Sciences at the Microscale and School of Physical Sciences, University of Science and Technology of China, Hefei 230026, China}
\affiliation{$^2$Shanghai Research Center for Quantum Science and CAS Center for Excellence in Quantum Information and Quantum Physics, University of Science and Technology of China, Shanghai 201315, China}
\affiliation{$^3$Hefei National Laboratory, University of Science and Technology of China, Hefei 230088, China}

\begin{abstract}
	The fermionic Hubbard model (FHM)\cite{hubbard1963proc}, despite its simple form, captures essential features of strongly correlated electron physics. Ultracold fermions in optical lattices\cite{esslinger2010fermi,bohrdt2021exploration} provide a clean and well-controlled platform for simulating FHM. Doping its antiferromagnetic ground state at half filling, various exotic phases are expected to arise in the FHM simulator, including stripe order\cite{zheng2017stripe}, pseudogap\cite{timusk1999pseudogap}, and $d$-wave superconductors\cite{xiang2022d}, offering valuable insights into high-temperature superconductivity\cite{anderson1987resonating,scalapino1986d,lee2006doping}. Although notable progress, such as the observation of antiferromagnetic correlations over short\cite{hart2015observation} and extended distances\cite{mazurenko2017cold}, has been obtained, the antiferromagnetic phase has yet to be realized due to the significant challenges of achieving low temperatures in a large and uniform quantum simulator. Here, we report the observation of the antiferromagnetic phase transition in a three-dimensional fermionic Hubbard system comprising lithium-6 atoms in a uniform optical lattice with approximately 800,000 sites. When the interaction strength, temperature, and doping concentration are finely tuned to approach their respective critical values, sharp increases in the spin structure factor (SSF) are observed. These observations can be well described by a power-law divergence, with a critical exponent of 1.396 from the Heisenberg universality class\cite{campostrini2002critical}. At half filling and with optimal interaction strength, the measured SSF reaches 123(8), signifying the establishment of an antiferromagnetic phase. Our results set the stage for exploring the low-temperature phase diagram of FHM.
\end{abstract}

\date{\today }
\maketitle

\section*{Introduction}

The Hamiltonian of the fermionic Hubbard model (FHM) takes a simple form, essentially consisting of two terms: one for nearest-neighbor hopping $t$ and the other for the on-site interaction $U$ between electrons with opposite spins. The FHM is capable of describing a wide range of strongly correlated electron physics\cite{arovas2022hubbard}, including interaction-driven metal-to-insulator transition\cite{jordens2008mott, schneider2008metallic}, quantum magnetism\cite{auerbach1998interacting}, and unconventional superconductivity\cite{xiang2022d,lee2006doping}. However, despite sixty years of intensive studies, an accurate understanding of its low-temperature physics remains elusive\cite{arovas2022hubbard,qin2022hubbard}. Exact analytical solutions are available only in one dimension (1D) or in the limit of infinite spatial dimensions\cite{giamarchi2003quantum,georges1996dynamical}. Although a broad variety of advanced numerical techniques have been developed, controlled quantitative studies are still scarce, particularly in the low-temperature regime. Quantum Monte Carlo (QMC) simulations typically suffer from the negative sign problem, while other methods, such as density matrix renormalization group\cite{Schollwoeck2005}, face limitations due to the exponential increase in computer memory requirements with system size. Even at half filling, simulations can become computationally prohibitive for sufficiently large systems at low temperatures\cite{qin2022hubbard}.

\begin{figure*}
	\centering
	\includegraphics[width = 0.75\linewidth]{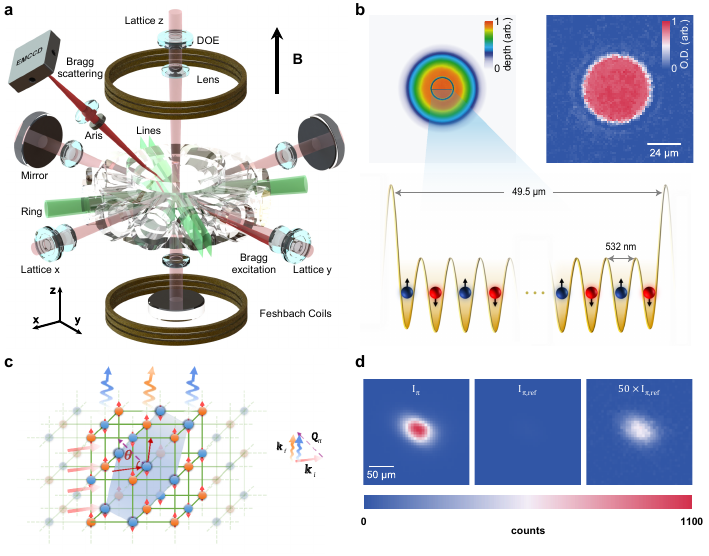}
	\caption{\textbf{Experimental scheme and setup}.
		\textbf{a}, Sketch of the experimental setup. Each lattice laser beam passes through a custom-designed diffractive optical element (DOE) and a focusing lens to achieve a flat-top intensity profile at the position of the atoms. After passing through a collimating lens, the lattice laser beam is retro-reflected by a mirror to create the standing wave fields. The cylindrical box trap is formed by a ring and two line beams, which isolate the central and, thus, the flattest region of the optical lattice. The Bragg laser is incident from the same vacuum window as the lattice laser $y$, providing precise irradiation of the atoms. The photons scattered from the atoms, following Bragg excitation, pass through an iris and a lens before being finally captured by an EMCCD.
		\textbf{b}, Illustration of the homogeneous optical lattice. The top-left image shows a contour plot of the trapping potential, which is a 2D cross-section perpendicular to the direction of the ring beam. The inner diameter of the dark ring measures approximately 49.5 $\upmu$m, determined by the repulsive box potential. The line is magnified in the bottom picture, indicating a 1D uniform optical lattice potential with a lattice spacing of 532 nm. In the top-right image, the {\it in situ} density distribution of atoms within the optical lattice at half filling is displayed, with the imaging plane being orthogonal to the direction of the ring beam.
		\textbf{c}, Principle of Bragg scattering. Here, $\theta$ represents the Bragg angle.
		\textbf{d}, Raw images of the Bragg signals progress from left to right, depicting $I_{\boldsymbol{\pi}}$, $I_{\boldsymbol{\pi},\text{ref}}$, with the signal amplified 50 times for $I_{\boldsymbol{\pi},\text{ref}}$.
	}
	\label{fig1}
\end{figure*}

In the past two decades, the development of modern laser and atomic techniques 
has led to the experimental realization of the FHM with ultracold atoms 
in optical lattices\cite{esslinger2010fermi,gross2017quantum}.
The precise control of Hamiltonian parameters has enabled these ultracold atomic systems to serve as powerful platforms for exploring the strongly correlated properties of FHM in low-temperature and doped regimes, which are not easily accessible through analytical and numerical methods\cite{bohrdt2021exploration}. To achieve this goal, a crucial and unavoidable step is the realization of the N\'eel phase transition and the antiferromagnetic phase in the FHM. This requires reaching low temperatures, preparing large and homogeneous systems, and developing new probing techniques. Along these lines, many important advances have been achieved. In 3D, the realizations of both Mott and band insulating phases\cite{jordens2008mott,schneider2008metallic}, as well as the observation of short-range quantum magnetism\cite{greif2013short}, have been reported. Notably, the spin structure factor (SSF) was observed to reach a value of $ S_{\boldsymbol{\pi}}\simeq 2$ at a temperature approximately $40\%$ higher than the N\'eel temperature $T_\text{N}$, suggesting the occurrence of short-range antiferromagnetic correlations\cite{hart2015observation}. In 2D, the advent of fermionic quantum gas microscopes\cite{gross2017quantum,gross2021quantum} has enabled the direct measurement of antiferromagnetic correlations as a function of distance across a square lattice of approximately 80 sites\cite{mazurenko2017cold}. At a temperature of $0.25t$, fitting the data with an exponentially decaying ansatz yielded a correlation length of $\xi=8.3(9)$, which approximately reaches the linear lattice size. Nonetheless, observation of critical phenomena, such as the power-law divergence of the SSF or the algebraic decay of the correlation function, has yet to be achieved as direct and conclusive evidence for the N\'eel phase transition.

In this work, we develop a large-scale quantum simulator for exploring the low-temperature physics of FHM by combining two key advancements: the generation of a low-temperature homogeneous Fermi gas in a box trap and the demonstration of a 3D flat-top optical lattice with uniform site potentials. At half filling, the Hamiltonian of the FHM exhibits $SU(2)$ symmetry, and, as the temperature $T$ decreases, it undergoes a transition to an antiferromagnetic N\'eel phase at the $T_\text{N}$, with the $SU(2)$ symmetry spontaneously breaking. As a consequence, it is believed that this antiferromagnetic phase transition shares the same universality class as the 3D classical Heisenberg model\cite{campostrini2002critical}, implying that the SSF should exhibit a divergent behavior, with $S_{\boldsymbol{\pi}} \propto (T-T_\text{N})^{-\gamma}$ and a critical exponent of $\gamma \simeq 1.396$. Away from half filling, the N\'eel temperature is significantly suppressed; nevertheless, the Heisenberg universality is expected to remain unchanged for small doping.
We measure the SSF of the realized FH system as a function of the interaction strength, temperature, and doping concentration, employing spin-sensitive Bragg diffraction of light\cite{hart2015observation}. Indeed, we do observe the expected critical divergent phenomena both at and away from half filling, providing conclusive evidence for the realization of the antiferromagnetic phase transition. Our results, particularly those away from half filling, are difficult to address even with state-of-the-art numerical computations, highlighting the advantages of quantum simulation and opening the avenue for tackling fundamental problems in strongly correlated Fermi systems\cite{norman2011challenge,keimer2015quantum}.

\section*{Experimental scheme and setup}

To investigate quantum magnetism within 3D FHM using ultracold Fermi gases, 
two critical conditions must be satisfied. 
First, it is essential to establish a spatially uniform 3D optical lattice, 
ensuring that the nearest-neighbor hopping $t$, 
on-site interaction $U$, and chemical potential $\mu$ 
remain nearly constant throughout the entire system or, 
at the very least, across a large portion of it. 
Second, it is imperative to reduce the system temperature $T$ 
below the exchange energy $J=4t^2/U$ for the establishment of 
long-range quantum magnetism. A common approach involves an initial 
preparation of degenerate Fermi gases using optical dipole traps, 
followed by slowly turning on the optical lattice potential. 
However, due to the nonuniform intensity distribution of the Gaussian laser beams, 
the resulting optical lattice potential also exhibits a Gaussian envelope\cite{Bloch2008}. 
This leads to variations in $t$, $U$, and $\mu$ as a function of spatial coordinate $\boldsymbol{r}$, resulting in the coexistence of multiple phases throughout the entire system. For instance, even with a moderate average atom number per lattice site, the system forms a multi-shell structure, changing from a band insulator to a metal, then to a Mott insulator, and finally back to the metal as one moves from the inner region outward\cite{Ho2009}. 
Moreover, the density distributions of Fermi gases in harmonic traps 
are inherently nonuniform, in contrast to the uniform density distributions of low-$T$ phases near half filling in the optical lattices. 
This leads to unavoidable heating effects due to the redistribution of the atomic density when loading Fermi gases into optical lattices. Additional compensating potentials have been employed to cancel out the underlying harmonic potential of the optical lattices\cite{hart2015observation,mazurenko2017cold}, effectively increasing the size of the uniform density region. Unfortunately, this approach does not rectifies the nonuniformity of $t$ and $U$, nor does it mitigate the heating induced by density redistribution during the lattice loading.

To overcome the aforementioned challenges, we present a novel approach that combines a box trap 
with a flat-top optical lattice (see Fig.~\ref{fig1}). 
This method involves loading a low-temperature homogeneous Fermi gas into the central uniform region, which is isolated by box trap\cite{navon2021quantum,li2022second}, within the flat-top optical lattice. Three key advantages are worth highlighting: (i) In the central region of the system, approximately $49.5~\upmu$m (equivalent to 93 lattice sites) in diameter, $t$, $U$, and $\mu$ are essentially independent of $\boldsymbol{r}$, despite the presence of residual potential disorder with a rms value of approximately 0.6\%. (ii) The density distribution of the homogeneous Fermi gas closely resembles that of strongly correlated lattice systems near half filling, effectively suppresses the density redistribution when turning on the lattice potential, resulting in near-adiabatic lattice loading. (iii) The average single-particle entropy of a homogeneous Fermi gas is approximately half of that of a nonuniform Fermi gas confined in a harmonic trap with the same reduced temperature $T/T_\text{F}$, enabling us to realize a 3D fermionic Hubbard (FH) system with very low single-particle entropy\cite{Ho2009}.

Our system consists of ultracold fermionic $^{6}$Li atoms, which are evenly populated among the hyperfine levels $|1\rangle \equiv | F=1/2, m_\text{F}=1/2 \rangle$ ($\left| \uparrow \right\rangle $) and $|3\rangle\equiv| F=3/2, m_F = -3/2 \rangle$ ($\left| \downarrow \right\rangle $), respectively. These atoms are confined within a cylindrical box trap with an inner diameter of 49.5~$\upmu$m and a height of 47.7~$\upmu$m at a magnetic field of 568~G, where the $s$-wave scattering length $a_{13}\simeq 0$. After devoting great technical efforts, we are able to prepare a non-interacting, homogeneous Fermi gas with a density of approximately $6.64\times 10^{12}/\text{cm}^3$ (corresponding to half filling in the optical lattice) and a low temperature of 0.041(1)$T_\text{F}$, or, equivalently, a low initial single-particle entropy $s$ of $0.216(2)k_{\text{B}}$, where $T_\text{F}$ is the Fermi temperature. Then, we gradually increase the lattice depth over a period of 18~ms to reach a final value of 6.25$E_\text{r}$, resulting in a nearest-neighbor hopping of $t/h=1.40(1)$~kHz. Here, $E_\text{r}/h=h/(8ma^2)=29.30$~kHz denotes the recoil energy, where $h$ is the Planck constant, $m$ represents the atomic mass, and $a=532$~nm is the lattice spacing. Simultaneously, we ramp up the magnetic field from 568~G to a predetermined final value within 12~ms, allowing us to achieve various on-site interactions using Feshbach resonance, with $U/h$ ranging from $6.00(2)$ to $28.69(9)$ kHz. In Fig.~\ref{fig1}b, we show a representative \textit{in situ} image of the realized low-temperature FH system, with $U/t\simeq 11$ and at half filling. It is observed that the atomic density distribution in the central region is highly uniform. In the vicinity of the system boundary, defined by the repulsive box potential with a finite width, the atomic density, or equivalently, lattice filling, experiences a rapid decrease over a distance of approximately 3~$\upmu$m. Utilizing available QMC data\cite{hart2015observation} at $U/t = 10$ and $T/t = 0.4$, along with the precise knowledge of the box trap potential, we employ the local density approximation (LDA) method to simulate the density distribution of atoms. The results suggest that approximately 68\% of the lattice sites are precisely singly occupied in the system. Moreover, the on-site interaction $U$, determined by the homogeneous lattice potential, remains nearly constant throughout the entire system.

To investigate the long-range antiferromagnetic order within the system, we employ the spin-sensitive coherent Bragg scattering technique, a method similar to that used in previous work\cite{hart2015observation}, as illustrated in Fig.~\ref{fig1}c. The incident Bragg photons, which are sensitive to spins and characterized by the wave vector $\boldsymbol{k_i}$, experience precisely opposite phase shifts when scattered by two different spin states. Hence, when the Bragg condition is fulfilled, $\boldsymbol{k_f}-\boldsymbol{k_i}=\boldsymbol{Q_\pi}$, with $\boldsymbol{Q_\pi}=\frac{2\pi}{a}(\frac{1}{2},\frac{1}{2},\frac{1}{2})$ representing a reciprocal lattice vector of the magnetic sublattice, the emitted photons with the wave vector $\boldsymbol{k_f}$ undergo constructive interference, leading to a coherent enhancement of the Bragg signal\cite{corcovilos2010detecting} $I_{\boldsymbol{\pi}}$:
\begin{equation}
	I_{\boldsymbol{\pi}}\propto\left[e^{-2W_{\boldsymbol{\pi}}}(S_{\boldsymbol{\pi}}-1)+ \frac{4\delta^2+s_0}{4\delta^2} \right], 
	\label{eq:detected intensity}
\end{equation}
where $\delta$ and $s_0$ are the dimensionless detuning and the on-resonance saturation parameter of the Bragg laser, respectively. Before the measurement, we linearly increase the lattice depth from $6.25E_{\text{r}}$ to $20E_{\text{r}}$ in 100~$\upmu$s to pin the optical lattice and greatly enhance the Debye-Waller factor $e^{-2W_{\boldsymbol{\pi}}}$. Subsequently, a 1.5~$\upmu$s Bragg pulse with $\delta\simeq13.7$ and $s_0\simeq30.6$ is applied, and the scattered photons in the $\boldsymbol{k_f}$ direction, within a full collection solid angle of $4\pi \times 8 \times 10^{-5}$ sr, are captured by an EMCCD. An additional reference image, denoted as $I_{\boldsymbol{\pi},\text{ref}}$, is acquired by turning off the optical lattice at 20$E_\text{r}$ and allowing a 50~$\upmu$s time-of-flight, during which the Debye-Waller factor $e^{-2W_{\boldsymbol{\pi,\text{ref}}}} $ approaches zero. These measurements enable us to access the spin structure factor $S_{\boldsymbol{\pi}}$, a crucial quantity for characterizing the long-range antiferromagnetic order\cite{corcovilos2010detecting}:
\begin{equation}
	\begin{aligned}
		S_{\boldsymbol{\pi}}&\equiv\frac{4}{N}\sum_{i,j} e^{i\boldsymbol{\pi}\cdot 	(\boldsymbol{R}_i-\boldsymbol{R}_j)} \left\langle \hat{S}^z_{i} \hat{S}^z_{j}\right\rangle\\
		&=e^{2W_{\boldsymbol{\pi}}} \left( 1+\frac{s_0}{4\delta^2}\right) \left( \frac{I_{\boldsymbol{\pi}}} {I_{\boldsymbol{\pi},\text{ref}}}-1 \right) +1,
		\label{eq:S}
	\end{aligned}
\end{equation}
where $N$ represents the total number of atoms. Here, the subscripts $i$ and $j$ denote lattice sites, $\hat{S}^z$ is a spin-$S$ operator along the spin-$z$ axis, and $\boldsymbol{R}$ is the site vector. The prefactor $\zeta=e^{2W_{\boldsymbol{\pi}}} \left( 1+\frac{s_0}{4\delta^2}\right)$ can be determined accurately. The obtained single-shot images are presented in Fig.~\ref{fig1}d, revealing that the Bragg signal's intensity is approximately one hundred times greater than that of the reference signal. This substantial increase signifies the successful establishment of a long-range antiferromagnetic order in the system.

\section*{Experimental results}

\begin{figure*}
	\centering
	\includegraphics[width = 0.8\linewidth]{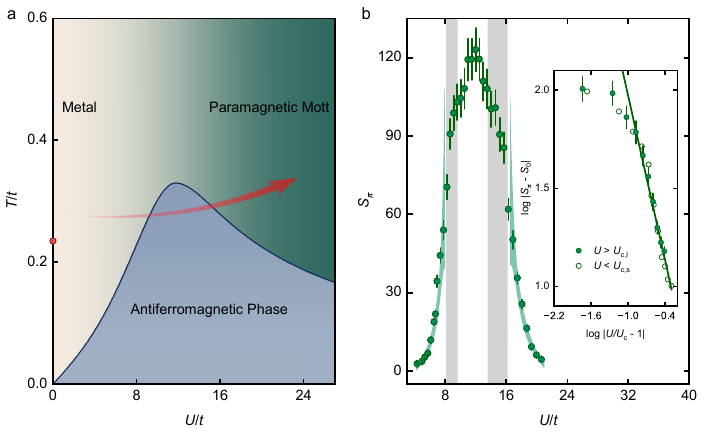}
	\caption{\textbf{Spin structure factor $S_{\boldsymbol{\pi}}$ as a function of $U/t$}. \textbf{a}, Schematic phase diagram of the 3D repulsive FHM at half filling. The red dot represents the calculated temperature at $U/t=0$, determined based on the measured temperature of a homogeneous Fermi gas, assuming an adiabatic raising of the optical lattice. The red curved band indicates the estimated temperature as a function of $U/t$. The discrepancy between the band and the point at $U/t=0$ arises from the slight non-adiabatic nature of the lattice loading process, which has been estimated through round-trip temperature measurements.  \textbf{b}, The measured $S_{\boldsymbol{\pi}}$ versus $U/t$. The experimental data (green dots) represent an average of approximately 125 independent measurements, with error bars indicating the standard errors.  We perform power-law fitting separately for data with $U/t < 8$ and $U/t > 16.4$, with the fitting formula being $S_{\boldsymbol{\pi}} - S_0 \propto \left| U/U_\text{c}-1 \right|^{-\gamma}$, where $\gamma=1.396$ is the critical exponent. The green and gray bands represent the 95\% confidence intervals for the fit of $S_{\boldsymbol{\pi}}$ and the fitted $U_\text{c}/t$, respectively, as determined by the fitting. The inset displays a log-log plot for $U < U_{\text{c,s}}$ (open circles) and $U > U_{\text{c,l}}$ (filled points). The solid line has a slope of $-1.396$.
	}
	\label{fig2}
\end{figure*}

We start with the half-filling case, where the physics of the FHM is well understood\cite{Staudt2000,Kozik2013}. As illustrated in Fig.~\ref{fig2}a, for small values of $U$, the system above $T_\text{N}$ behaves as a metallic paramagnet, and the N\'eel transition arises from a spin-density wave (SDW) instability with a small modulation of sublattice magnetization\cite{Hirsch1987}. The mean-field theory predicts an exponentially small N\'eel temperature $T_{\text{N}} \sim e^{-t/\rho_0 U}$, where $\rho_0$ represents the density of states at the Fermi surface\cite{Hirsch1987}. At strong interaction strengths, the paramagnet above $T_\text{N}$ evolves into a Mott insulator. The low-energy sector of the FH Hamiltonian reduces to the Heisenberg model, and the antiferromagnetic phase transition occurs at\cite{domb2000phase} $T_\text{N} \approx 0.9575 J$. Between these two limiting branches lies a smooth crossover region, where $T_\text{N}$ reaches its maximum, denoted as $T_{\text{N, max}}$, at an optimal $U$. While the dynamical mean-field theory\cite{georges1996dynamical} suggests $T_{\text{N, max}} \approx 0.45t$ at $U \approx 10t$, numerical simulations indicate a lower value\cite{Kozik2013, qin2022hubbard}, $T_{\text{N, max}} \approx 0.33t$, within the range of $8t\le U \le 9t$. Note that, we depict $T_{\text{N, max}}$ at a higher on-site interaction of $U\sim 11.5t$ in Fig.~\ref{fig2}a for two reasons: First, near the system boundary, the atomic density becomes suppressed and deviates from half filling, which is more pronounced for smaller values of $U$. Consequently, due to the sensitivity of the SDW instability on the Fermi surface, it is expected that the $T_\text{N}$ for the small-$U$ branch will be reduced. Second, the residual disorder in the lattice potential\cite{paiva2015cooling}, characterized by a randomness strength $\varrho \sim t$, results in a modified exchange energy\cite{duan2003controlling} $J'=4t^2U/(U^2-\varrho^2)$. This might lead to a slight increase in $T_\text{N}$ for the Heisenberg branch. Therefore, the optimal $U$ is expected to be larger than the numerical estimate.

To explore the phase diagram, we prepare the half-filled FH system with various on-site interactions. At $U=0$, the lowest attainable temperature, $T\simeq0.21t$, can be calculated by assuming an adiabatic evolution from the homogeneous Fermi gas to the FH system (red dot in Fig.~\ref{fig2}a). However, due to the non-adiabatic nature of the lattice loading, a slight temperature increment of approximately 20\% is observed. As $U$ increases, the thermometry of the FH system becomes extremely challenging. Therefore, the temperature dependence on $U$ can only be estimated, as indicated by the red arrow in Fig.~\ref{fig2}a. The measured SSF as a function of on-site interactions is displayed in Fig.~\ref{fig2}b. Starting with a small value of $S_{\boldsymbol{\pi}}=2.8(2)$ at $U\simeq4.31t$, the SSF increases smoothly as $U$ increases until reaching $U \simeq7t$. It then enters into a sharp increase until $U \simeq 9t$, and finally slows down and reaches a maximal average value $S_{\boldsymbol{\pi}}= 123(8)$ at $U \simeq 11.96t$. Notably, in a single measurement, $S_{\boldsymbol{\pi}}>300$ can even be observed due to critical fluctuations. When $U$ is further increased, $S_{\boldsymbol{\pi}}$ becomes suppressed, undergoes a rapid drop around $U\simeq16t$, and then decreases to $4.4(2)$ at $U\simeq 20.62t$. The curve of $S_{\boldsymbol{\pi}}$ is roughly symmetric around $U \simeq 11.75t$. 

The antiferromagnetic phase transition in the half-filled FHM can be classified into the universality class of the 3D Heisenberg model, owing to the inherent $SU(2)$ symmetry in its Hamiltonian\cite{campostrini2002critical}. Consequently, as an infinitely large and fully equilibrated FH system approaches the critical point, the correlation length increases as $\xi \sim |{\bf x}| ^{-\nu}$, and the SSF diverges as $S_{\boldsymbol{\pi}} \sim \xi^{\gamma/\nu} \sim |\bf x|^{-\gamma}$, with $\nu=0.712$ and $\gamma=1.396$ being the universal critical exponents\cite{campostrini2002critical}. The symbol ${\bf x}$ represents the proximity to the critical point in the parameter space of interaction, temperature, and atomic density. We find that the $S_{\boldsymbol{\pi}}$ data for $U/t \leq 8$ and $U/t\geq 16$ are well captured by $S_{\boldsymbol{\pi}}-S_0 \propto \left| U_{\text{c}} / U-1 \right| ^{-\gamma}$, where $S_0$ denotes the background contribution. The determined N\'eel transition points are $U_{\text{c,s}} =8.84(33)$ and $U_{\text{c,l}} =14.86(42)$ for the small-$U$ and large-$U$ branches, respectively, with $\gamma=1.396$ being fixed. The inset of Fig.~\ref{fig2}b clearly demonstrates the scaling phenomenon, where the data coincide with a straight line having a slope of $-1.396$, signifying the power-law divergence. Remarkably, the $S_{\boldsymbol{\pi}}$ data for the small-$U$ and large-$U$ branches collapse approximately onto a single curve, providing a direct illustration for the universality. 

In practical scenarios, it is anticipated that the correlation length $\xi$ will eventually saturate at a finite $\xi_{\rm sat}$. This saturation is likely to occur on a scale smaller than the linear system size, primarily for the following reasons. As the critical point is approached, following the Kibble-Zurek mechanism, the relaxation time diverges as $\xi^z$, with the dynamic exponent $z \gtrsim 2$, making it challenging for the system to reach equilibrium. Additionally, the experimental system is susceptible to imperfections. The atomic density is suppressed near the boundary, leading to a deviation from half filling. Furthermore, even with a small strength, the residual disorder in the lattice potential significantly reduces the correlation length. As a result, the divergence of the SSF is attenuated and saturated at $\xi_{\rm sat}^{\gamma/\nu}$, as manifested by the crest curve within the range $9<U/t<15$. Moreover, the obtained SSF, which is an average of $S_{\boldsymbol{q}}$ around $\bm{q}\simeq\bm{\pi}$ in the Brillouin zone, is significantly lower than the actual SSF of the system. This discrepancy is especially pronounced in the antiferromagnetic phase, where $S_{\boldsymbol{q}}$ approximates a delta function centered at $\bm{q}=\bm{\pi}$. Therefore, even a slight momentum mismatch in the Bragg scattering measurement would result in a dramatic decrease in the measured $S_{\boldsymbol{\pi}}$ value.

\begin{figure*}
	\centering
	\includegraphics[width = \linewidth]{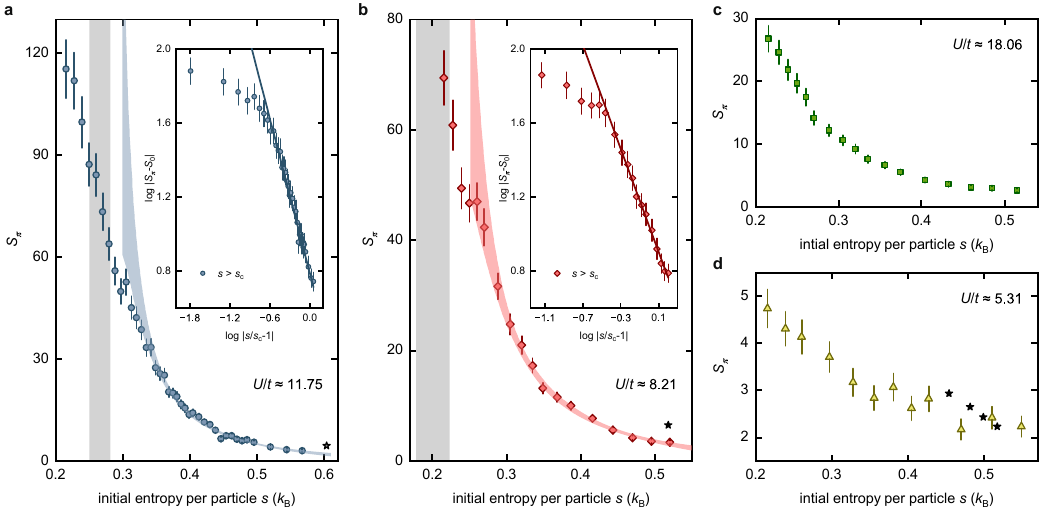}
	\caption{\textbf{Spin structure factor $S_{\boldsymbol{\pi}}$ as a function of initial entropy per particle $s$}.
		\textbf{a-d}, Experimental results for four different values of $U/t$, specifically $11.75$, $8.21$, $18.06$, and $5.31$, respectively. Each data point represents an average based on approximately 125 independent measurements, with the standard statistical error calculated. Here, $s$ refers to the initial entropy per particle in the case of non-interacting homogeneous Fermi gases. We fit the data with $s > 0.336k_\text{B}$ and $s > 0.275k_\text{B}$ to the power-law scaling function $S_{\boldsymbol{\pi}} - S_0 \propto \left| s/s_\text{c}-1\right|^{-\gamma}$ for $U/t \simeq 11.75$ and $8.21$, respectively, where $\gamma=1.396$ is the critical exponent. The blue (red) and gray bands in \textbf{a} (\textbf{b}) depict the 95\% confidence intervals for the fit of $S_{\boldsymbol{\pi}}$ and the fitted $s_\text{c}$, respectively, as determined by the fittings. The insets in \textbf{a} and \textbf{b} are log-log plots of the data, and the solid lines have a slope of $-1.396$. The black stars in \textbf{a}, \textbf{b}, and \textbf{c} are the numerical results (see text).
	}
	\label{fig3}
\end{figure*}

We then measure $S_{\boldsymbol{\pi}}$ as a function of temperature $T$ and present the results in Fig.~\ref{fig3}a-d for $U/t\simeq11.75$, $8.21$, $18.06$, and $5.31$, respectively. It is worth noting that, instead of using the temperature of the FH system, we employ the initial entropy per particle $s$, determined from the non-interacting, homogeneous Fermi gas. This choice is necessitated by the extremely difficulty in accurately measuring temperature in the low-entropy regime of the FH system. Fortunately, as long as $T$ exhibits a smooth dependence on $s$, the critical divergence is expected to remain unchanged, as clearly evidenced in Figs.~\ref{fig3}a and b. For $U\simeq11.75t$ and $s \geq 0.336k_\text{B}$, the power-law fit $S_{\boldsymbol{\pi}}-S_0 \propto \left| s/s_{\text{c}}-1\right| ^{-\gamma}$ yields a critical entropy of $s_{\text{c}}=0.27(1)k_\text{B}$. Similarly, for $U\simeq8.21t$ and $s \geq 0.275k_\text{B}$, the fitting results in $s_{\text{c}}=0.20(1)k_\text{B}$. In contrast, for $U\simeq 18.06t$ the critical scaling appears to be relatively weak, indicating that our system is in proximity to but has not yet reached the N\'eel transition. Finally, for $U\simeq5.31t$, the critical scaling is hardly discernible, suggesting that $T_\text{N}$ would occur at a significantly lower temperature. 

With the measured potential shape of the box trap, initial entropy per particle, and the QMC data\cite{hart2015observation}, we apply the LDA to calculate the $S_{\boldsymbol{\pi}}$ values. It is important to note that QMC data are not available for strong interaction strengths and low temperatures, especially away from half filling. Therefore, LDA results are only obtained at high temperatures and for $U/t\simeq 5.31$, $8.21$, and $11.75$, which are represented by the black stars in Figs.~\ref{fig3}a, b, and d. For $U\simeq5.31t$, the LDA values of $S_{\boldsymbol{\pi}}$ agree well with the experimental measurements. However, for $U\simeq8.21t$ and $11.75t$, the numerical calculations are approximately 20\% greater than the experimental results. While the measured $S_{\boldsymbol{\pi}}$ are indeed lower than their actual values, there are two additional reasons contributing to the numerical-experimental discrepancies: (i) The entropy per particle of the FH system is slightly higher than that of the homogeneous Fermi gas (i.e., $s$), due to the inherent non-adiabaticity during the lattice loading. (ii) The numerical calculations have not taken into account the residual disorder in the lattice potential. 

\begin{figure*}
	\centering
	\includegraphics[width = 0.8\linewidth]{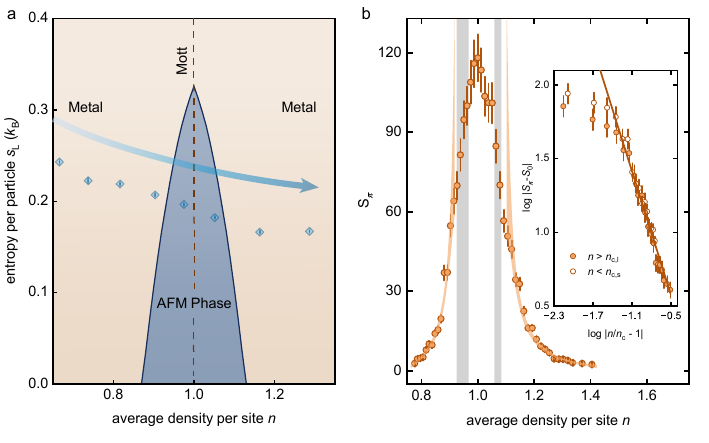}
	\caption{\textbf{Spin structure factor $S_{\boldsymbol{\pi}}$ as a function of average density per site $n$}.
		\textbf{a}, Schematic phase diagram of the doped FHM at $U/t \simeq 11.75$. The blue dots represent the measured initial entropy per particle $s$ for non-interacting homogeneous Fermi gases at  various values of $n$. The error bars indicate one standard deviation. The blue curved band represents the estimated entropy per particle, $s_{\text{L}}$, within the optical lattice. 
		\textbf{b}, The measured $S_{\boldsymbol{\pi}}$ versus $n$. Each data point and error bar are the mean and standard error of the mean, respectively, calculated from approximately 125 independent measurements. The data for $n<0.915$ and $n>1.108$ are separately fitted to the power-law scaling function $S_{\boldsymbol{\pi}} - S_0 \propto \left| n/n_\text{c}-1\right|^{-\gamma}$, with the critical exponent $\gamma$ fixed at 1.396. The orange and gray bands indicate the $95\%$ confidence intervals for the fit of $S_{\boldsymbol{\pi}}$ and the fitted $n_\text{c}$, respectively. The inset displays a log-log plot for $n < n_{\text{c,s}}$ (open circles) and $n > n_{\text{c,l}}$ (filled points). The solid line has a slope of $-1.396$.
	}
	\label{fig4}
\end{figure*}

We now consider the 3D FH model away from half filling. In the doped regime, simulation time of QMC increases exponentially with system size and inverse temperature, primarily due to the fermionic sign problem, leaving little conclusive knowledge available. Nevertheless, second-order perturbation theory suggests that, at a fixed $U$, the N\'eel temperature is significantly reduced by doping. Recently, state-of-the-art theoretical calculations using the dynamical vertex approximation (D$\Gamma$A)\cite{schafer2017interplay} or diagrammatic Monte Carlo\cite{Lenihan2022} predict that with 10\%-20\% doping, $T_\text{N}$ reaches zero at a quantum critical point (QCP)\cite{schafer2017interplay}. Furthermore, even at doping levels of 5\%-15\%, the ground state transitions from an antiferromagnet to an incommensurate SDW state, mainly driven by the doping-induced deformation of the Fermi surface. The transition is considered to be in the 3D Heisenberg universality class for small doping, but it remains largely unknown for large doping, where several closely competing low-energy states may exist, and the QCP can play an essential role\cite{schafer2017interplay}.

In the experiment, we prepare the FH system at the optimal $U \simeq 11.75t$ and lowest temperature, where $S_{\boldsymbol{\pi}}$ reaches its maximal value at half filling. The phase diagram is depicted in Fig.\ref{fig4}a, where the vertical axis represents the entropy per particle $s_\textsc{L}$ of the FH system. The blue dots and the thick arrow represent the measured $s$ and the estimated $s_\text{L}$ as a function of the average density per lattice site $n$, respectively, where $n=1$ corresponds to half filling. The discrepancy between $s_\text{L}$ and $s$ indicates a non-adiabatic increase in entropy. The experimental results for $S_{\boldsymbol{\pi}}$ are displayed in Fig.\ref{fig4}b as a function of $n$, clearly showing sharp increases or decreases near the critical points. The power-law scaling function $S_{\boldsymbol{\pi}}-S_0 \propto |n/n_{\text{c}}-1|^{-\gamma}$, with $\gamma=1.396$, fits the experimental data well for $n<0.91$ or $n>1.11$, resulting in critical values of $n_{\text{c,s}}=0.95(1)$ and $n_{\text{c,l}}=1.07(1)$, respectively. Furthermore, in the log-log plot, similar to the case of varying interaction strength or temperature, the data points for $n<0.91$ or $n>1.11$ collapse onto a solid line with a slope of $-1.396$. While our experimental data are not sufficient to provide a precise determination of the magnetic critical exponent $\gamma$, the quality of the fit suggests consistency with the 3D Heisenberg universality.

\section*{Outlook}

We have developed an advanced quantum simulation platform for the 3D FHM by first preparing a low-temperature homogeneous Fermi gas in a box trap and then adiabatically loading it into a 3D optical lattice with uniform site potentials. This approach ensures nearly uniform on-site interaction $U$ and nearest-neighbor hopping $t$ across the lattice, which consists of approximately 800,000 sites and operates at a temperature well below $T_\text{N}$. We have observed the critical divergences in the SSF, which is smoking-gun evidence of the antiferromagnetic phase transition. Our setup offers great promise for exploring the low-temperature and doped physics of FHM\cite{keimer2015quantum}. It enables us to observe and investigate exotic quantum phases in 3D or quasi-2D by tuning the interlayer coupling strengths. Examples include spin-charge separation\cite{vijayan2020time,senaratne2022spin} in higher dimensions, pseudogap\cite{timusk1999pseudogap} in repulsive settings, and stripe order\cite{zheng2017stripe} at specific fillings, by utilizing techniques such as spin-sensitive Bragg diffraction of light\cite{hart2015observation}, momentum-resolved microwave spectroscopy\cite{li2023observation,stewart2008using}, and Bragg scattering in the long-wavelength limit\cite{li2022second,senaratne2022spin}. Moreover, by adding optical superlattices to the setup, we can realize coupled plaquette arrays and explore the possible $d$-wave pairing therein\cite{Ying2014}. These quantum simulations will provide valuable experimental data, enhancing our understanding on the role of quantum magnetism in the mechanism of high-temperature superconductivity. Furthermore, by tuning the atomic interactions to be attractive, we aim to realize the long-sought single-band superfluid in the attractive FHM\cite{Hofstetter2002,hartke2023direct} and explore the underlying physics of the BCS-BEC crossover\cite{chen2005bcs,zwerger2011bcs} within a lattice context.

We thank J.-P. Hu, J. Schmiedmayer, Y. Qi, Y.-Y. He, and Q.-J. Chen for discussions. This work is supported by the Innovation Program for Quantum Science and Technology (Grant No. 2021ZD0301900), NSFC of China (Grant No. 11874340), the Chinese Academy of Sciences (CAS), the Anhui Initiative in Quantum Information Technologies, the Shanghai Municipal Science and Technology Major Project (Grant No.2019SHZDZX01), and the New Cornerstone Science Foundation.

\bibliography{references}

\end{document}